\documentclass[fleqn,usenatbib]{mnras}
\usepackage{newtxtext,newtxmath}

\usepackage[T1]{fontenc}

\usepackage{graphicx}	

\newcommand{\Msun}{\mbox{\,$M_{\odot}$}}        
\newcommand{\Zsun}{\mbox{\,$Z_{\odot}$}}        

\title[Maximum Black Hole mass across Cosmic Time]{Maximum Black Hole mass across Cosmic Time}
\author[J. S. Vink et al.]
{Jorick S. Vink\thanks{E-mail: jorick.vink@armagh.ac.uk}, 
Erin R. Higgins, 
Andreas A. C. Sander, 
Gautham N. Sabhahit\\
\\
$^{1}$Armagh Observatory and Planetarium, College Hill, Armagh, BT61 9DG, Northern Ireland
}

\date{Accepted 2021 March 16. Received 2021 February 1; in original form 2020 December 30}

\pubyear{2020}

\begin{document}
\label{firstpage}
\pagerange{\pageref{firstpage}--\pageref{lastpage}}
\maketitle

\begin{abstract}
 At the end of its life, a very massive star is expected to collapse into a black hole. 
  The recent detection of an 85\,\Msun\ black hole from the gravitational wave event GW\,190521 appears to present a fundamental problem as to how such heavy black holes exist above the approximately 50\,\Msun\ pair-instability limit where stars are expected to be blown to pieces with no remnant left. 
 Using MESA, we show that for stellar models with non-extreme assumptions, 90..100\,\Msun\ stars at reduced metallicity ($Z/\Zsun \leq 0.1$) can produce blue supergiant progenitors with core masses sufficiently small to remain below the fundamental pair-instability limit, yet at the same time lose an amount of mass via stellar winds that is small enough to end up in the range of an "impossible" 85\,\Msun\ black hole.
The two key points are the proper consideration of core overshooting and stellar wind physics with an improved scaling of mass loss with iron (Fe) contents characteristic for the host galaxy metallicity. Our modelling provides a robust scenario that not only doubles the maximum black hole mass set by pair instability, but also allows us to probe the maximum stellar black hole mass as a function of metallicity and Cosmic time in a physically sound framework.
\end{abstract}

\begin{keywords}
gravitational waves -- stars: black holes -- stars: evolution -- stars: massive -- stars: mass-loss -- stars: winds, outflows
\end{keywords}



\section{Introduction}

The direct detection of the first gravitational waves from the merger of two heavy black holes (BHs) in GW\,150914 confirmed one of the toughest predictions of Einstein's Theory of general relativity. But while satisfying the world of Physics in general, for Astrophysics this was only the beginning: many were surprised by the large BH masses of respectively $36$ and $29$\,\Msun\ \citep{abbott16}, showcasing how the new field of multi-messenger Astrophysics had just re-opened the field of stellar evolution in a spectacular fashion. Stellar-mass black holes had previously been revealed by their interaction in binary systems \citep{orosz11}, but the maximum stellar BH mass in our Milky Way is not higher than roughly 15..20\,\Msun\ \citep{belc10}. While we know that very massive stars (VMS) above 100\,\Msun\ exist \citep{crowther2010,vink2015}, this mass is significantly diminished via stellar winds already during core hydrogen (H) burning \citep{vg12}. The heavy Nature of the black hole, as  measured by LIGO/VIRGO therefore supported the assumption that the gravitational wave event occurred in a part of the Universe still pristine in its enrichment with heavy elements (`metallicity ($Z$)'), lowering stellar wind mass loss \citep{vink2001,vdek05}. A low-$Z$ solution was widely accepted until the announcement of a 70\,\Msun\ black hole in LB-1 \citep{liu19}, spurring stellar evolution theorists to avoid heavy mass loss in the Milky Way \citep{belc20,groh20}, either by arbitrarily lowering the mass-loss rates of VMS  -- seemingly contradicting VMS mass-loss calibrations \citep{vg12} -- or by invoking the presence of a strong dipolar surface magnetic field that could quench the wind \citep{petit17}. While such magnetic fields in some 5-10\% of massive OB stars do indeed exist, no B-fields have yet been detected in VMS \citep{bagnulo20}. The problem of the formation of a $70\,M_\odot$ BH in a solar-metallicity environment apparently resolved itself when the spectral signatures of LB-1 were re-interpreted \citep{abdul20,el-baldry20}. 

The recent discovery of GW\,190521, involving the merger of a $85^{+21}_{-14}$\,\Msun\ and a $66^{+17}_{-18}$\,\Msun\ BH \citep{abbott20}, was not only record-breaking in terms of obtained BH masses, but also represents an exciting challenge. The masses of both black holes in GW\,190521 are at the limit of what is called the second\footnote{The first mass gap refers to a observational gap between the most massive neutron stars and `lightest' black holes.} mass gap between approx.\ 50-130\,\Msun, where stars cannot collapse into BHs due to pair instability resulting from electron-positron pair production \citep{fowler64}. 

Beside a regime where the whole star is disrupted by a so-called pair-instability supernova (PISN) \citep{barkat67,glatzel85,fryer01,umeda02,scannapieco05,langer07,kasen11,kozy17,mapelli20}, there is also a regime where electron-positron pair production does not disrupt the star as whole, but causes significant violent pulses leading to enhanced mass loss before an eventual `pulsational pair-instability supernova' (PPISN) \citep{chatz12,chen14,woosley17,marchant19,leung19}. In contrast to PISNe, PPISNe are leading to a BH as there is a remaining iron core that collapses. Still, the pulses before the eventual collapse remove so much mass, that the combination of PISNe and PPISNe leads to a significant `forbidden' mass regime, where no first-generation heavy black hole should be found. The lower boundary of this regime is commonly considered to be approximately 50\,\Msun\ \citep{farmer19,woosley20}, as the hydrogen (H) envelope is generally assumed to be lost, either through binary interaction or through strong mass loss in the most massive stars.

Given inherent uncertainties of both the BH masses and the second mass gap, it is mainly the $85\,M_\odot$ BH in GW\,190521 that surprised the Astronomical community and led to the speculation that such heavy BHs of up to 85\,\Msun\ are most likely `second generation' BHs, implying they must have merged from lighter BHs in an earlier event \citep{abbott20}. The preferred solutions involve mergers of lower-mass BHs or stars in dense cluster/galactic environments \citep{fragione20,romero-shaw20}, however as {\it both} BHs in the system are above the 50\,\Msun\ boundary, this would imply an arguably contrived situation involving at least 2 double mergers, i.e.\ at least involving 4 objects.
While one cannot rule out such a scenario, we will show that the formation of heavy BHs on the order of $85\,M_\odot$ neither requires an earlier BH merger nor more exotic scenarios such as a modified gravitation theory \citep{moffat20}. The aim of our paper is {\it not} to reproduce all aspects of GW\,190521, but instead to critically assess core overshooting and mass-loss assumptions and to check if an 85\,\Msun\ BH may be able to form under reasonable conditions.  We will indeed reveal that our uncertain understanding of the evolution of (very) massive stars due to our limited knowledge of wind mass loss has lead to an underestimation of the lower boundary of the second mass gap at low metallicity ($Z$). 

In the following, we will show that some blue supergiants (BSGs) at low, but not necessarily zero metallicity ($Z$), on the order of 90..100\,\Msun\ can retain most of their H-rich envelopes and avoid the pair-instability regime. By critically assessing stellar wind mass loss we present a solution for what has been considered an "impossible"\footnote{We place the word "impossible" between inverted commas as there is no fundamental physical limit being broken.} BH mass in a wide range of host environments.

\section{Evolution of very massive stars}

In this paper, we show that BSGs of order 90..100\,\Msun\ can silently collapse to BHs. The prime reason this solution was not considered until the discovery of GW\,190521 was that previous authors generally assumed the H-envelope to be lost, either due to binary Roche-Lobe overflow or strong mass loss in an individual star, e.g. as a Luminous Blue Variable (LBV). However, as the currently most robust estimates of close binarity in massive stars are of order 50\% \citep{sana13} and some of those will already merge on the H-burning main sequence, only a fraction of massive stars are expected to be subjected to Roche-Lobe overflow. Furthermore, while the most massive single stars may loose significant amounts of mass as LBVs at high $Z$, there is little evidence this would be the case at lower $Z$, In fact, the incidence of LBVs at low $Z$ seems to suggest a drop in LBV phenomenology, on both empirical and theoretical grounds \citep{kalari18,vink18bsj,grassitelli20}. 

In short, there is no a priori reason to assume that VMS loose their H-envelopes, as was for instance the case in a recent extensive stellar evolutionary study (also with MESA) on PISNe and PPISNe, resulting in a lower boundary of the second mass gap of $\simeq$50\,\Msun\ \citep{farmer19}.
This particular study was based on model grids leading to high precision and small error bars. In fact it was even argued that mass loss was one of the smaller uncertainties and that the $^{12}\mathrm{C}(\alpha,\gamma)^{16}\mathrm{O}$ nuclear reaction was the larger contributor to the error budget. However, as we will argue below, such small error bars may not be realistic if the assumption of the removal of the H-envelope is questionable. In our study, we do not aim for high precision, but we aim for accuracy, at least in the first instance, while precision studies are left for future works.

The aim of our paper is to show that VMS of order 90..100\,\Msun\ can lose little mass at low $Z$ to still have a sufficient total mass reservoir available to produce a first generation BH of order 85\,\Msun. At the same time, the models need to avoid too large carbon-oxygen (CO) core masses, as otherwise pair production during oxygen (O) burning might produce pulses that could remove too much mass.

The most straightforward way to remain below this critical boundary is to start off with an initial mass ($M_{\rm init}$) that is comparatively low (of order 90..100\,\Msun), as more massive stars have progressively larger convective cores \citep{yusof13, koehler2015, vink2015}. Furthermore, the fractional increase of the core by convective overshooting must remain relatively low in order to maintain a compact core, such as values found by \citet{ekstroem2012, higgins2020} of order $\alpha_{\rm ov}$ $\simeq$0.1. Note that we do {\it not} argue that {\it all} massive stars should necessarily have small cores, as there is plentiful evidence from astroseismology \citep{bowman2020}, eclipsing binaries \citep{higgins19}, and the width of the main sequence \citep{vink10,brott11} that many massive stars have larger cores, implying some form of extra-mixing, commonly attributed to "overshooting". 
In this study, we show that small amounts of overshooting are needed to produce BSG progenitors that may retain a large fraction of their H envelope. Stars with larger amounts of overshooting would evolve to the red supergiant (RSG) phase where strong mass loss would decrease the mass of the H-envelope. 
Astroseismology results in the mass range up to approx. 25\,\Msun\ suggest overshooting values in terms of step overshooting $\alpha_{\rm ov}$ from values close to 0 up to 0.44 \citep{bowman2020}, with an equal if not slightly higher chance for low overshooting values of order $\alpha_{\rm ov}$ $\simeq$ 0.1 used for our BSG models than higher $\alpha_{\rm ov}$ cases. Given that what we are interested in here in our study is the {\it Maximum} black hole mass as a function of $Z$, we can be confident that this is set by the {\it low} overshooting, small core models, which we explore in the following.

 \begin{table}
     \centering
     \begin{tabular}{c c c c c c c c c c }
        \hline
         Model & $Z$/\Zsun & $M_{\rm{init}}$ & $M_{\rm{f}}$ & $M_{\rm{env}}$ & $M_{\rm{He}}$ & $M_{\rm{CO}}$ &  \\
        \hline\hline
        A1 & $0.1$  & $90$  &  $80.4$ & $42.9$ & $37.5$ & $32.7$ & cc \\
        A1-Alt & $0.1$  & $90$  &  $74.0$ & $35.2$ & $38.8$ & $33.8$ & cc \\
        A2 & $0.1$  & $110$ &  -- & -- & $45.1$ & $39.5$ & PI \\
        B1 & $0.01$ & $90$  &  $87.0$ & $46.1$ & $40.9$ & $34.6$  & cc \\
        B2 & $0.01$ & $110$  &  -- & -- & $52.8$ & $45.6$ &  PI \\
        \hline

    \end{tabular}   
        \caption{Fundamental parameters of our representative models, including the initial metallicity $Z$ and mass $M_\mathrm{init}$. For all models, we also list the total mass $M_\mathrm{f}$, envelope mass $M_\mathrm{env}$, and CO core mass $M_\mathrm{CO}$ of the final model stage. Models which are subject to pair instability (PI) were stopped at this point.}

    \label{tab:masses}
\end{table}

\section{MESA modelling}

After describing our evolution strategy modelling, we utilise MESA (v12115 \cite{paxton2011,paxton2013,paxton2015,paxton2018,paxton19}) and update its physics as described below.
We evolve the objects at least through core O-burning and we check if the models encounter pair instability (PI). Successful models for heavy BH formation thus require two key criteria. The first one is that of a CO core mass that remains below 37\,\Msun, the second criterion is that of a massive envelope (above about 40\,\Msun). 

A representative set of models is listed in Table 1. 
The range of initial masses is comprised of $90$-$110\,M_\odot$ with initial metallicities of $1$-$10\%\,Z_\odot$. The initial composition of each model encompasses a scaled-solar metallicity based on the abundances by \citet{GS98} with $Y = 0.266$ and $Z_\odot = 0.017$. This scaling is also applied to the employed OPAL opacity tables from \citet{RogersNayfonov02}. 

The standard mixing length theory of convection by \citet{BohmVitense58} is included with $\alpha_{\rm{MLT}}$ $=$ 1.82 \citep{choi2016}. The Ledoux criterion is employed with standard semiconvective mixing described by an efficiency parameter $\alpha_{\mathrm{sc}}= 1$. We incorporate extra mixing by convective core overshooting via the `exponential-overshooting' method. 
This overshooting description assumes an exponential mixing profile that decays as a function of distance from the core. The extent of this mixing region is defined by the parameter $f_{\mathrm{ov}}$, that gives a measure of inverse of the slope of the exponential decay. The diffusion coefficient in our models decreases from $D_0$ near the core ($f_0 = 0.001$ times the pressure scale height below the core) until a minimum coefficient of $D_{\rm{min}} = 0.01$ is reached. 
Overshooting dredges extra H from the envelope into the core during H-burning, meaning that the helium core will be larger with increased $f_{\mathrm{ov}}$. 
In our models, we use $f_{\mathrm{ov}}$ $=$ 0.01 (equivalent to step-overshooting $\alpha_{\mathrm{ov}}$ $=$ 0.1). 
While this treatment is common in the  evolution models of massive stars \citep{higgins2020, ekstroem2012}, it is a crucial parameter deciding stars become blue or red supergiants. 
According to recent calculations \citep{farmer19}, a CO core above $\sim 37\,M_\odot$ (corresponding to a He core of $\sim 50\,M_\odot$) would lead to pair instability. Our models yielding massive BHs have CO cores masses below this limit (cf.\ Table 1), but could enter this regime for considerably larger efficiencies of overshooting. 

The MLT$^{++}$ routine in MESA has been used to evolve stars with very high initial masses through their post-main sequence stages by reducing the superadiabaticity in the radiatively dominated envelopes.
Our models use the standard MLT$^{++}$ values set in MESA. 
MLT$^{++}$ essentially reduces the contribution of radiative energy transport and offers an alternative, more efficient energy transport mechanism in these regions. MLT$^{++}$ is activated in all regions where the superadiabaticity is greater than $f_1 = 10^{-4}$, irrespective of the fraction of energy transported by radiation $(\lambda_1 = -1 )$. The actual reduction in superadiabaticity is determined by the parameter $f_2 = 10^{-2}$, with smaller values resulting in larger reduction in superadiabaticity. 
Although MLT$^{++}$ has a physical significance, its usage here is a numerical necessity for such high mass models to evolve to late burning stages without technical problems, e.g.\ due to extremely short time-steps \citep{paxton2013}. 

The effects of rotation have been included with the transport of angular momentum and chemical elements treated as diffusive processes through rotationally-induced instabilities. These instabilities include dynamical and secular shear instabilities as well as the Solberg-Høiland, and Goldreich-Schubert-Fricke instabilities, and Eddington-Sweet circulation. The effects of a Spruit-Taylor dynamo have not been included in our models, allowing for differential rotation rather than solid-body rotation. The efficiencies of these instabilities are included as described by \citet{heger2000}. For all models, we implement a rotation rate as a function of the critical velocity for each mass with $\Omega / \Omega_{\mathrm{crit}}= 0.2$, corresponding to rotation rates of $\sim$ 150\,km\,s$^{-1}$. Since the effects of rotation lead to larger core sizes with increased $v_{\rm{rot}}$, a slower rotation is preferred to keep the core compact and reach the maximum BH mass. Faster rotating stars will likely form lower-mass BHs as the increased rotation will not only increase the core size, but beyond a certain limit also promote bluewards evolution towards a Wolf-Rayet (WR) star, which will lose a significant portion of the total mass by the time of core collapse. 

All models are calculated with the `standard' time-step resolution of 10$^{-4}$ and are evolved until core collapse unless convergence issues occur at earlier stages. In all cases, our models have been evolved to core O-burning, where models either reach the pair instability criteria or continue O-burning until a BH is formed. In MESA, models that undergo pulsations at the onset of core O exhaustion due to pair production prompt a condition when $\gamma$ -- summed over all cells -- falls below 4/3.
MESA models of massive stars undergoing pulsational pair instability during O burning have been previously studied in \citet{farmer19}, although in the context of pure helium stars. The presence of a massive envelope in our models ($> 35 M_\odot$) during O burning (e.g.\ in our model A1) helps to avoid this $\gamma < 4/3$ regime. 

\begin{figure*}
    \includegraphics[width=\textwidth]{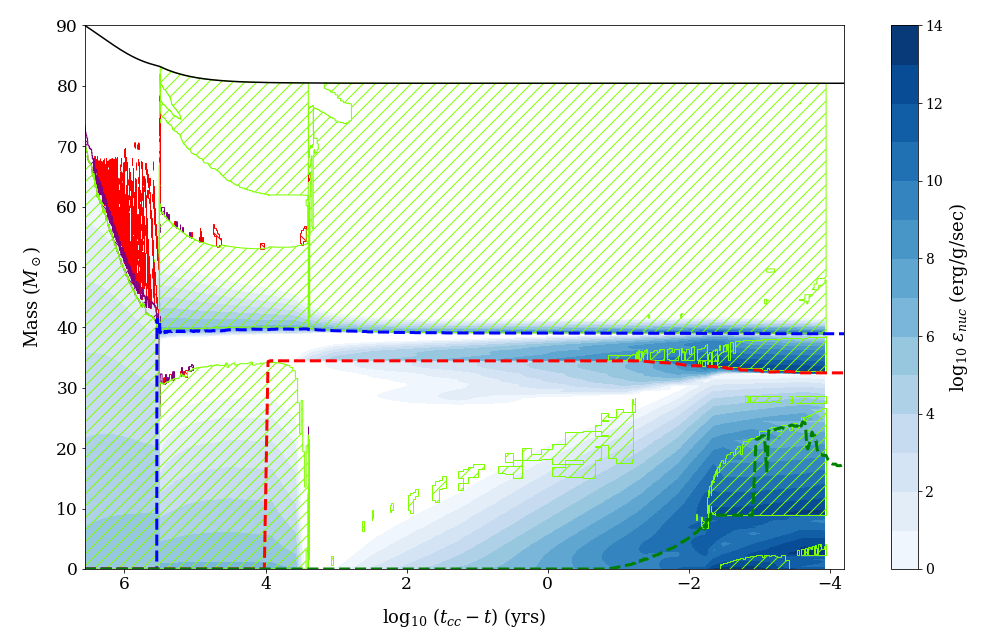}
    \caption{Evolution of the internal structure of our model A1 with initial mass 90\Msun\ and metallicity 0.1\Zsun, as a function of mass (\Msun) and time until core collapse. The color bar represents the nuclear burning regions. The green hatched region denotes the convective zones, purple region denotes the small overshooting layer above the H-burning core and the red region denotes layers with semi-convective mixing. The dashed blue and red lines represent the He and CO core boundaries respectively.}
    \label{fig:model1Kipp}
\end{figure*}

Figure \ref{fig:model1Kipp} showcases the stellar structure of our model A1 (a $90\,M_\odot$ star at $0.1\,\Zsun$) as a function of evolutionary timescale (`Kippenhahn diagram').
For the purpose of calculating our stellar models within the scope of this study, we have adopted the standard treatments of convection and rotation in stellar evolution, as well as the common sets for atomic data and nuclear reactions. 
While we have assessed the parameter space around our adopted inputs, we find that our conclusions are not impacted by the implementation of convection parameters or the necessary treatment of the technical aspects, such as the use of MLT$^{++}$ to run stellar models with high initial masses to late evolutionary stages. 

In contrast, choices which affect the physics determining the eventual CO core size and the loss of envelope mass, are very important for our results. Therefore, we have tested the extent of core overshooting during core H-burning, rotation rates, and the implementation of mass loss through stellar winds. We find a strong dependence on the amount of overshooting, such that a small fractional increase is preferable to maintain a compact core. Similarly, a modest rotation rate is required to keep the core compact and avoid evolution towards homogeneously mixed stars. Consequently, we estimate that our scenarios for forming massive BHs just below the PI limit are applicable for stars with initial surface rotation velocities of $\sim$50-200kms$^{-1}$, which is representative for the bulk of observed massive stars at low metallicity \citep{RamirezAgudelo2013}.

\section{Improved treatment of wind mass loss} 

As stellar mass is the prime quantity for stellar evolution, the loss of mass due to stellar winds is a decisive factor in massive star evolution. The rate of mass loss is a complex function of stellar properties such as the stellar effective temperature that changes during the evolutionary track. 
Therefore, evolution models rely on physical theoretical or empirical descriptions for specific temperature ($T_\mathrm{eff}$) regimes. The application of such formulae beyond the validity regime where they were established has to be carefully considered as inadequate extrapolations may lead to over- or underestimations by orders of magnitude \citep{sv20}. 

A standard approach inherent to state-of-the-art stellar evolution codes is the collection of various mass-loss treatments. In MESA, the generally applied formula for massive stars is the so-called `Dutch wind scheme'. For hot stars on the main sequence, the models rely on line-driven Monte Carlo radiative transfer calculations \citep{vink2001}, while a different relation involving dust-driving is employed for cooler star winds \citep{deJager1988}. For stars which have their outer H envelope removed, a third formula for Helium stars is employed. In MESA and other state-of-art codes this final formalization is generally based on an empirical study of WR stars in the Milky Way \citep{nl2000}. This final treatment does not accurately reflect the wind physics at the low metallicities considered in the context of massive black holes as GW progenitors.  

In our stellar evolution models discussed below, those that include our improved wind treatment, such as A1 and B1, are able to maintain their H envelopes intact. Those with 
the standard MESA treatment, such as model A1-Alt, lose extra mass when the model switches either to the RSG recipe (as is the case for A1-Alt) or to the WR recipe as was the case for some other test models. Either way, in this MESA "Dutch" standard treatment, there is rapid envelope loss during core He-burning, even at low $Z$. 

The reason for the strong mass loss in model A1-alt is the early switch from a hot-star wind treatment \citep{vink2001} to the one for cool stars \citep{deJager1988}. 
The mass-loss description for the cool star regime is employed to describe dusty RSGs, which have temperatures of just $3000$-$4000$\,K. While the switch in the standard treatment occurs already at $10 000$\,K, the BSGs in this temperature range ($8\,000$-$12\,000$\,K) are too hot to form dust and their winds are expected to be driven by iron-dominated gas opacities rather than dust. Consequentially,
the utilization of the treatment of dust-driven winds leads to an overestimation of the actual mass loss, and our modelling would not be able to make BHs more massive than 75\,\Msun. 

When we employ the more physical line-driven wind description in this regime for stars above the so-called second bi-stability jump above 8000 K \citep{petrov16}, all our model stars are BSGs, avoiding the regime of the RSGs, and can retain a sufficiently large amount of their H envelope to make BHs of order 80-87\,\Msun\ at low $Z$ (see models A1 and B1).

As was already alluded to, another aspect inherent to evolution modelling with codes such as MESA is the employment of the third formula. Intended to describe the mass loss of He-enriched WR stars \citep{nl2000}, it is applied to evolved stars above $10\,000$\,K. 
As the mass loss scales with iron content, but not with metals that can be produced via self-enrichment such as CNO \citep{vdek05,sander2020}, we apply the physically motivated standard hot star wind treatment \citep{vink2001} also for our evolved models as a WR-like treatment would yet again lead to artificially high mass-loss rates at low $Z$. 

\subsection{Potential extra Mass loss near the Eddington Limit}

A potential concern for the proposed heavy BH scenario would occur if the suggested wind treatment would underestimate mass-loss rates in close proximity to the Eddington limit \citep{vink2011} or even above the Eddington limit due to continuum driving \citep{shaviv2000}. 
The first question concerns mass-loss during core H burning as this is where the stars spend most of their lifetime. Electron scattering values during core H burning for model A1 are only of the order of 0.3 which ensures that the metallicity dependent rates of \citet{vink2001} are safely applicable. The next potential hurdle could occur during the approximately 10 times shorter core He burning phase. In this phase the electron scattering Eddington parameter climbs to 0.7, which is marginally larger than the values that were employed for the derivation of the \citep{vink2001} parametrisation, but checking more recent Monte Carlo computations by  \citep{vink18sms} show that the mass-loss rates stay comfortably low (of order 10$^{-5}$\,\Msun/yr). At very low $Z$ additional metal line opacities would not change this assertion, but for our case we tested the "worst case" scenario of the total Eddington value for model A1 at 10\% the solar metallicity using the sophisticated 1D PoWR stellar atmosphere code \citep{sander2020,Graefener2002}, which provides the most accurate (flux-mean) opacities in 1 dimension.

From the analysis of the PoWR atmosphere models, we conclude that while the metallicity-dependent continuum opacity is larger than that of just electron scattering, the values remain sub-Eddington, and a continuum driven wind is not launched during core-He burning. Metallicity-dependent line driving needs to be carefully analysed in order to assess if the predicted mass-loss rates from Vink et al. (2001) are significantly underestimated in this regime. If mass-loss rates are as high as 10$^{-4}$\,\Msun/yr, of order ten solar masses could potentially be lost during core He burning. However, mass-loss rate predictions in this regime are highly uncertain and still in its infancy, and it would be equally or more likely that the true mass-loss rates are lower than 10$^{-4}$\,\Msun/yr in which case a negligible amount of mass would be lost during core He burning. As the key physics involve metallicity-dependent driving, even in the worst case scenario of a high mass-loss rate, our proposed scenario would still be valid, but merely the critical metallicity, where we have a steep drop in the maximum BH mass, would shift towards lower $Z$ by approximately a factor of two. 

Note that in reality, atmospheres in close proximity to the Eddington limit might either inflate \citep{Graefener2012} or become porous \citep{shaviv2000}. In such cases, 3D hydrodynamical modelling \citep{Jiang+2018} would become necessary. Future 3D radiative driving and hydrodymical stellar models \citep{grassitelli20} are needed for accurate physical modelling, as current 3D hydrodynamical models still rely on approximate Rosseland mean opacities. 

While our detailed atmosphere models show that we do not reach a regime of continuum-driven winds during core-He burning, this could happen in the very last stages of evolution. Here, the excess luminosity is so high that super-Eddington mass-loss is likely. However, the timescales for this mass-loss are low, thus leading to modest time-averaged mass-loss rates. For numerical reasons, it is a standard procedure in stellar evolution modelling to not include explicit mass loss during late burning stages. However, we can do a common posterior estimation \citep{Renzo2020} by assuming that all the luminosity that exceeds the Eddington limit can be used to remove mass that leaves the star with a velocity similar to the escape speed. Integrated over the whole time after core-He burning, this leads to a loss of $0.19\,M_\odot$. This value is comparable to the amount of mass lost during core collapse and negligible in the context of our overall scenario.

\section{Results}

\begin{figure*}
    \includegraphics[width=\textwidth]{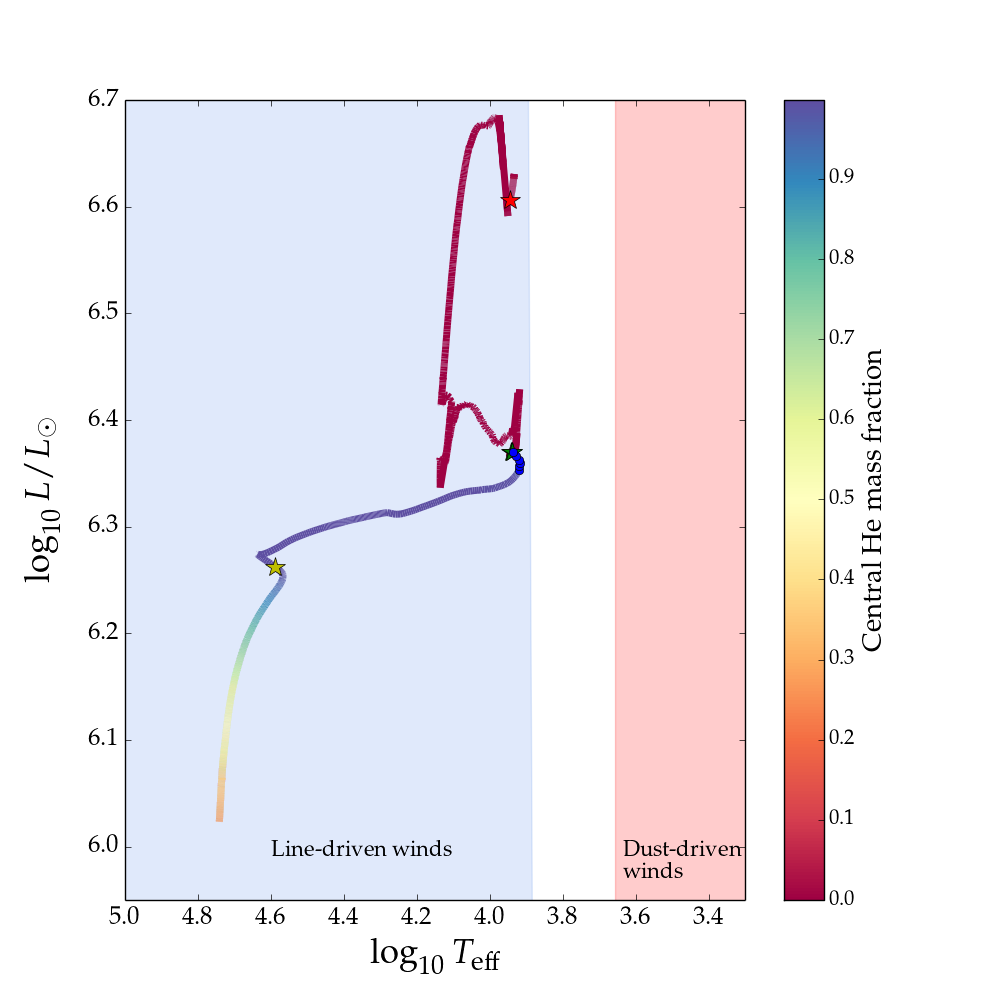}
    \caption{Evolution of model A1 in a Hertzsprung-Russell diagram. The colour bar represents the core He abundance, with a yellow star showing the TAMS position, a blue star illustrating the end of core He-burning, and a red star marking the end of core O-burning. Blue dots (near the blue star) show time-steps of 50,000 years after core H exhaustion, where time is spent as a BSG (i.e. above log T$_{\rm {eff}}$ $>$ 3.65). Shaded regions highlight the area in the HRD where RSGs (red) evolve with dust-driven winds or BSGs (blue) evolve with line-driven winds.} 
    \label{fig:model1HRD}
\end{figure*}

Model A1, a $90\,M_\odot$ star at $0.1\,Z_\odot$, becomes a BSG around 10,000 K (see the HRD of Fig.\,\ref{fig:model1HRD}) and eventually collapses into a BH of 80\,\Msun. Note that this model evolves through all stages up to Si burning with a final CO core mass of $32.7\,M_\odot$. At this point, the star retains a massive, H-rich envelope of $42.9\,M_\odot$, which is even larger than the CO core. Contrary to common visualisations of evolutionary tracks of VMSs where phases beyond core He-burning are usually omitted, we do show the full evolution in Fig.\, \ref{fig:model1HRD}, including the shorter O-burning phase where the HRD positions are less certain, see also \cite[e.g.][]{Renzo2020, belc20}. 

In contrast to recent calculations determining the maximum BH mass below the PI gap \citep{farmer19,woosley20}, we do not a priori assume that stars forming heavy BHs must have lost their outer envelope. Instead, we have critically assessed the treatment of mass loss for different temperatures employed in massive star evolution codes and improved their physical treatment, scaling the mass-loss rate with iron (Fe) which is the host galaxy metallicity.

Generally for massive stars below the Humphreys-Davidson limit, stellar evolution models encounter significant mass loss during the helium (He) burning phase as cooler red supergiants (RSGs). 
To simulate this, we also evolved a star with the same properties as model A1 but now employing the standard mass-loss treatment with higher rates. This model (A1-Alt) dips into the RSG regime where its mass-loss rate increases by more than an order of magnitude. This leads to higher mass loss during He burning, and although the model collapses to a BH, its final mass is never found to be above 75\,\Msun. 
For higher mass initial masses (e.g. 110\,\Msun, Model A2), the resulting BH mass does not grow either as the resulting PI would either remove the mass in pulsations or destroy the star completely. We do not follow these pulsations as this is not the focus of our study, but stop our calculations entering the PI regime.

\subsection{The role of metallicity}

As a next step, we explore the maximum BH mass across Cosmic time by assessing the role of initial host $Z$.  
Models B1 and B2 have one order of magnitude lower $Z$ than models A1 and A2. Model B1 with an initial 90\,\Msun\ now forms an 87\,\Msun\ BH, while the 110\,\Msun\ model B2 still reaches the PI regime due to its higher CO core. 
Our metallicity exploration thus tells us robustly that the resulting maximum BH mass is approximately 90..100\,\Msun, which is in line with the masses inferred for all currently known GW events, including GW\,190521, as well as marginal triggers, which are GW candidates with signal-to-noise ratios just below the standard threshold for detection \citep{udall2020}.

As mentioned earlier, we do not claim that {\it all} 90..100\,\Msun\ stars at low $Z$ produce such heavy BHs. Statistics from GW events seem to indicate a break in the BH mass distribution around 40\Msun\ \citep{fishbach20,abbott21} and \cite{abbott21} find that only $\sim$3\% of systems have primary masses greater than 45\Msun. This is partly expected to be a result of the $Z$ dependent upper BH mass that we infer in this paper, but we also wish to remind the reader that even independent of the $Z$ argument, some massive stars will interact with a binary companion, and as also discussed earlier there is evidence for higher values of core-overshooting for several massive stars. In other words, our low-$Z$ "small core" solution for an 85\Msun\ BH only covers a part of the parameter space (at the maximum), and further parameter studies are needed for inferences on BH populations.

\begin{figure*}
    \includegraphics[width=\textwidth]{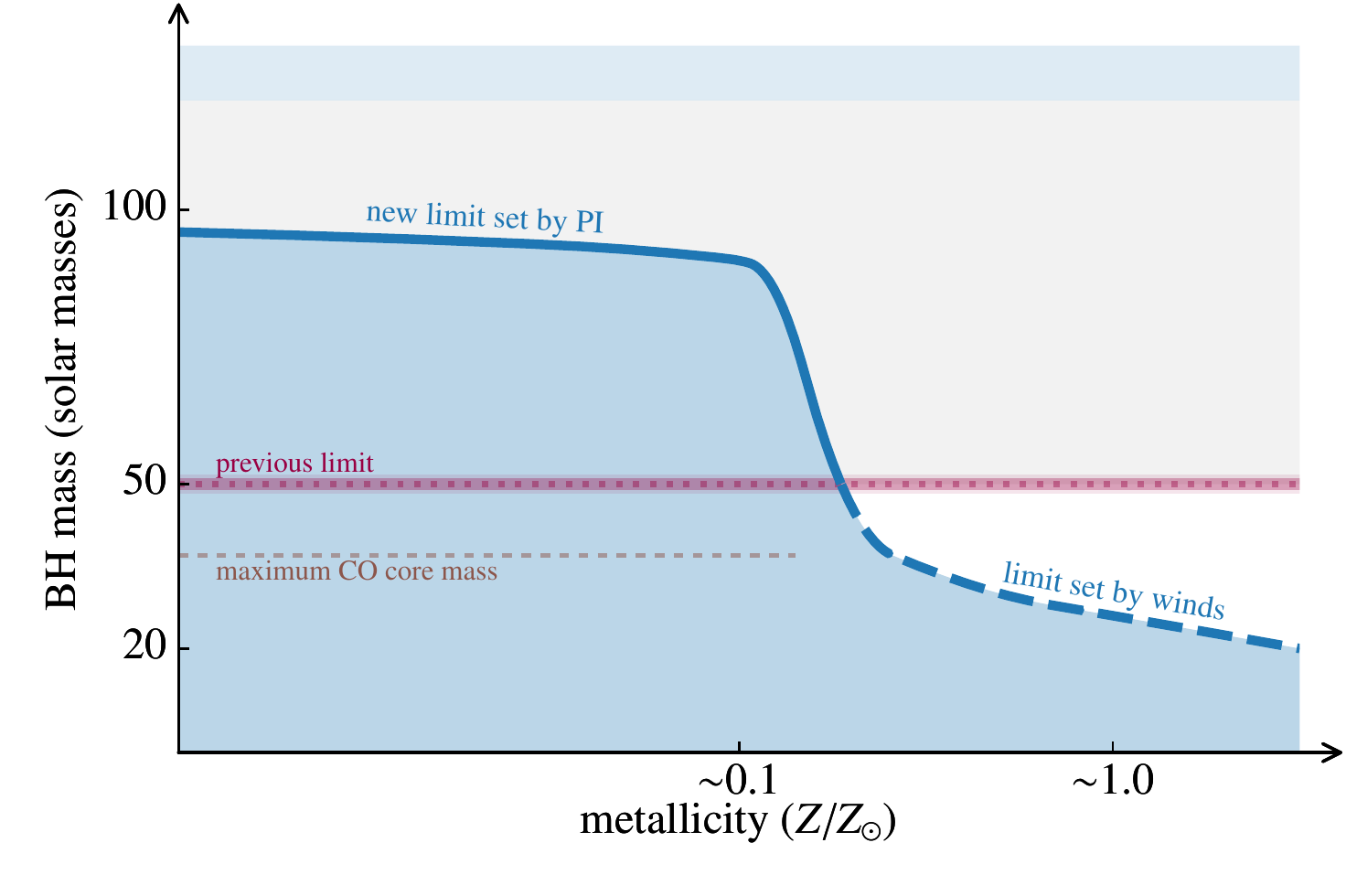}
    \caption{Schematic view of the maximum black hole mass below the pair instability gap as a function of metallicity: 
    at low $Z$, the maximum BH mass (blue solid line) is set by PI, where above this line the PI gap (grey region) is shown. Above a threshold metallicity, the maximum BH mass is determined by $Z$-dependent stellar winds (blue dashed line). 
    The red line sets the previous maximum BH mass below the PI gap, while the light blue region at the top of the diagram provides the upper limit of the PI gap.} 
    \label{fig:maxBHscheme}
\end{figure*}

A summarizing sketch of our findings is given in Fig.\,\ref{fig:maxBHscheme}. The darkest blue shaded area shows the range of BH masses as a function of $Z$. The blue line represents the maximum BH mass. For low $Z$ stars, the limit is set by PI where BHs cannot exist (illustrated by the grey region).
At a threshold metallicity, the maximum BH mass steeply declines due to the removal of the envelope by $Z$-dependent winds. At high $Z$, the maximum BH mass drops below the limit set by PI, but is now set by the further removal of layers via $Z$-dependent wind mass-loss.

The red solid line outlines the previous PI limit of order $50\,M_\odot$ set by the maximum CO core mass (of $\sim 37\,M_\odot$) along with a diminished envelope.
As we illustrate in Fig.\,\ref{fig:maxBHscheme}, the infall of the envelope allows the BH to effectively {\it double} the previously obtained mass limits \citep{woosley17,farmer19} which were based on absent or significantly decreased envelopes. 
A pivotal outcome of our study is that the maximum {\it initial} mass to avoid pulsations does not significantly change with metallicity below a certain threshold $Z$, compared to the amount of retained envelope mass which changes drastically with metallicity. 

\subsection{The silent collapse of the star}

To actually identify the final masses of our evolutionary models with
the resulting BH masses, the core collapse of the star should not be
accompanied by a supernova explosion as this might lead to a
considerable loss of mass. Indeed, massive stars below the (pulsational)
PI regime are most likely to produce an immediate black hole -- a
so-called `failed supernova' -- if their core is compact enough
\citep{oconnor2011,ertl2016}\footnote{See also \cite{gilkis16} and \cite{burrows20} for alternatives to 'failed' supernovae and compactness respectively. For cases where rotation induces more mass loss in the collapsar see also \citet{zhang08,nagataki07,tominaga09,woosley12}}. This is fulfilled for all of our models
avoiding pair instability (A1, A1-Alt, B1). Still, to reach the high BH
masses observed in recent GW events, also the outer layers of the star
need to fall into the newly formed BH. To avoid a significant ejection
of mass from the outer layers at core collapse, thus also the envelope
needs to be sufficiently compact.

To determine whether our stars modelled in the evolutionary calculations
undergo a silent collapse including also the envelope, we study the
so-called `compactness' of the core and the envelope. A common
measurement for the core compactness \citep{oconnor2011,ertl2016} is the
parameter
\begin{equation}
  \xi_\mathrm{2.5} = \frac{2.5}{r(M = 2.5 M_\odot)
[1000\,\mathrm{km}]}\mathrm{.}
\end{equation}
Our models at $0.1\,Z_\odot$ avoiding pair instability have a core
compactness of $\xi_\mathrm{2.5} = 0.77$ (models A1 and A1-Alt). While
the lower boundary of $\xi_\mathrm{2.5}$ for a failed supernova is still
a matter of debate \citep{ertl2016}, this value is considerably above
all estimates for what is considered to be able to explode as a
supernova \citep{ertl2020}. Moreover, with CO core masses of over
$30\,M_\odot$, all of our models remaining below the pair instability
gap are in a regime where recent studies expect them implode directly at
the end of their life \citep{ertl2020,woosley20}.
until Fe core due to time step problems.

To check, what fraction of the envelope falls into the newly formed
black hole, we consider the envelope compactness parameter
\citep{fernandez18}
\begin{equation}
  \xi_\mathrm{env} = \frac{M_\mathrm{f} [M_\odot]}{R_\mathrm{f} [M_\odot]}
\end{equation}
with $M_\mathrm{f}$ and $R_\mathrm{f}$ denoting the total mass and
radius of the star at core collapse. At the time of core collapse, our
model stars will appear as blue supergiants with $\xi_\mathrm{env} =
0.1$. While not as compact as WR stars, this is an order of magnitude
higher than for red supergiants, meaning that the amount of ejected mass
is fairly low. Based on recent simulations for ejecta masses in failed
supernovae \citep{fernandez18}, our models with $\xi_\mathrm{env} = 0.1$
will lose less than $0.2\,M_\odot$ as failed supernovae only yield large
ejecta masses for red supergiants where the outer layers contain a
considerable amount of mass and are very weakly bound to the star at
all. Hence, even at $0.1 Z_\odot$ our evolution model A1 is able to
produce a black hole of about $80\,M_\odot$.

\section{Summary \& Outlook}

In our pilot study, we find BH masses in the range of $80$-$90\,M_\odot$ for $Z$ of 10\% of the solar value and below. The blue progenitor stars of these BHs retain most of their H envelope due to the inefficiency of line-driven wind mass loss at these low, but non-extreme metallicities. 
Low metallicity galaxies of order 10\% the solar value, such as Sextans A, are observed in the Local Universe, and thus there is no need to invoke Population {\sc iii}-type environments for creating heavy BHs.
Despite their massive envelope, our model stars never become RSGs, preventing the occurrence of dust-driven mass loss. We do not consider any binary interactions in our study, but we note that the smaller radial extent of our blue supergiants limits the interaction frequency with potential companions. For more discussion on binary interactions and Pop {\sc iii} stars, see e.g.\ \citet{tanikawa20,umeda20,farrell21}.

While full grids of stellar evolution models would be necessary to pin
down the precise physical limits and exact $Z$-dependent boundaries between BH
formation and pair instability, we can confidently predict the maximum BH mass below the PI gap to be on the order of $90$-$100$\,\Msun\ at low $Z$ (below 10\% solar $Z$). Our predictions on the maximum BH mass are based on evolution models with $M_\mathrm{init} \sim 90$-$110\,M_{\odot}$ in the $0.01$ - $0.1\,Z_\odot$ range. At the highest considered metallicity, we find an optimum situation producing an $80\,M_\odot$ BH while avoiding pair instability, allowing us to constrain the upper stellar mass limit as well as upper $Z$-limit for \textit{first generation} BH formation. At larger $Z$, enhanced wind mass loss will rapidly drop the maximum BH limit down to the established value of about $50\,$\Msun. With the occurrence of WR-type mass loss, which is a strong function of luminosity-to-mass-ratio and $Z$ \citep{sander2020,sv20}, the effective BH mass limit can shrink even further \citep{woosley20}. As sketched in Fig.\,\ref{fig:maxBHscheme}, the strong mass loss of massive, envelope-stripped stars during He burning can remove a considerable amount of mass, which further needs to be quantified in dedicated follow-up studies.

Our physically motivated evolution modelling highlights that first generation black holes can only be ruled out for a small range of masses between approximately $100$ and $130\,M_\odot$. 
The heavy black holes obtained in recent GW events such as GW 190521 show that wind mass loss at low $Z$ is a crucial ingredient that needs to be carefully considered in stellar evolution and population synthesis modelling to avoid blurring our perception of how the Universe evolved into what we see today. From the analysis and modelling performed here, we may already confidently conclude that for low-$Z$ host galaxies it is possible to create {\it first generation} BHs up to values as large as $\sim$ 90\Msun, without the need to invoke {\it second generation} BH formation, extreme assumptions, or exotic physics.

\section*{Acknowledgments}
We warmly thank the MESA developers for making their stellar evolution code publicly available, and the anonymous referee for their detailed and constructive comments.
JSV and ERH are supported by STFC funding under grant number ST/R000565/1.
AACS is \"{O}pik Research Fellow.

\section*{Data Availability}

The data underlying this article will be shared on reasonable request
to the corresponding author.



\bibliographystyle{mnras}
\bibliography{references}

\bsp	
\label{lastpage}

\end{document}